\documentstyle[aps]{revtex}

\begin{document}

\draft

\title{Parton Interaction Rates in the Quark-Gluon Plasma}
\author{M.H. Thoma}
\address{Institut f\"ur Theoretische Physik, Universit\"at Giessen,\\
35392 Giessen, Germany}
\date{\today}
\maketitle

\begin {abstract}

The transport interaction rates of elastic scattering processes of
thermal partons in the quark-gluon plasma are calculated beyond the
leading logarithm approximation using the effective perturbation
theory for QCD at finite temperatures developed by Braaten and
Pisarski. The results for the ordinary and transport interaction
rates obtained from the effective perturbation theory are compared
to perturbative approximations based on an infrared cut-off by the
Debye screening mass. The relevance of those interaction rates for a
quark-gluon plasma possibly formed in ultrarelativistic heavy ion
collisions are discussed.

\end{abstract}



\section{Introduction}

Interaction or damping rates of parton scattering processes in a
thermalized quark-gluon plasma (QGP) are of great physical
significance. For example, the inverse rates related to the elastic
scattering of thermal partons ($gg\rightarrow gg$, $gq\rightarrow
gq$, $qq\rightarrow qq$) give the mean free paths and typical
interaction times i.e., relaxation times used in the collision term
of the Boltzmann equation. Hence they can be used for an estimate of
the thermalization time of a pre-equilibrium parton gas in
ultrarelativistic heavy ion collisions \cite{r1} and the maintenance
of the thermal equilibrium by comparing the interaction rates versus
the cooling rate. Of course thermalization times calculated in this
way at finite temperature should be valid only for situations which
do not start too far from equilibrium. On the other hand, this
thermalization time does not depend on the somewhat ambiguous
definition of the beginning and the end of the pre-equilibrium stage
as the one deduced from numerical studies of ultrarelativistic heavy
ion collisions (HIJING \cite{r2}, parton cascade \cite{r3}). After
all, both methods lead to similar results, namely a fast
thermalization of the gluons i.e., an isotropic and exponential
momentum distribution, after about 0.2 fm/c for LHC and 0.3 fm/c for
RHIC \cite{r4}. The full equilibration of the phase space density can
be investigated, on the other hand, using the inelastic interaction
rates e.g., $gg\rightarrow ggg$ and $gg\rightarrow q\bar q$,
describing the chemical equilibration process of the QGP \cite{r4}.

Furthermore the elastic rates are the basic inputs for the energy
loss of a parton in the QGP \cite{r5,r6,r7,r8,r9,r10,r11,r12} and the
viscosity of the QGP \cite{r13}. The first quantity is related to a
possible signature of the QGP in ultrarelativistic heavy ion
collisions (jet quenching) \cite{r7,r14,r15}, while the latter
provides informations about the role of dissipation in the
hydrodynamic expansion phase \cite{r16}.

Finally, damping rates are especially suited for studying problems of
quantum field theory at finite temperature like gauge dependences and
infrared singularities in perturbation theory \cite{r17}. In
particular, the puzzle of the gauge dependence of the gluon damping
rate at rest (plasmon damping) in naive perturbation theory started a
recent developement in finite temperature field theory, which lead to
a powerful effective perturbation theory \cite{r18}. Assuming the
weak coupling limit ($g\ll 1$) effective propagators and vertices are
obtained by resumming self energies and vertices in the high
temperature limit (hard thermal loops), which are used for soft
external momenta of the order $gT$, while bare Green's functions are
sufficient for hard momenta of the order $T$. In this way gauge
independent results for physical quantities are found, which are
complete to leading order in the coupling constant. (The gauge
independence of this method is still under discussion by evaluating
the damping rates in a general covariant gauge. However, using an
appropriate infrared regularization for the gauge dependent part,
independence of the gauge fixing parameter can be shown
\cite{r19,r20,r21,r22}.) In addition, screening effects are included
yielding an improved infrared behavior. In summary, the
Braaten-Pisarski method provides a consistent treatment for
quantities which are sensitive to the momentum scale $gT$, meaning a
crucial improvement compared to the naive perturbation theory at
finite temperature.

Here, we will present calculations of the interaction rates based on
the Braaten-Pisarski technique and study their physical relevance for
the QGP. Two different kinds of interaction rates are considered. In
the next section we will discuss the ordinary interaction rate
$\Gamma $ -- in the following simply called interaction rate -- of a
parton with hard momentum, defined by $\Gamma \equiv n\; \sigma $,
where $n$ is the particle density of the QGP and $\sigma $ the cross
section of the scattering process under consideration. In naive
perturbation theory without resummation $\Gamma \sim \alpha _s^2\, T$
is found ($\alpha _s =g^2/4\pi$). However, $\Gamma $ turns out to be
quadratically infrared divergent. Using the Braaten-Pisarski method,
$\Gamma \sim \alpha _s\, T\, \ln(1/\alpha _s)$ follows, which is only
logarithmically infrared divergent
\cite{r7,r22,r23,r24,r25,r26,r27,r28,r29}. The interaction rate is
enhanced by a factor $1/\alpha _s$ compared to the one obtained from
naive perturbation theory due to the infrared regularization inherent
in the resummation method. The logarithmic term, reflecting the
logarithmic infrared divergence, arises from the sensitivity of the
rate to the momentum scale $g^2T$, which cannot be treated within the
Braaten-Pisarski method.

In section 3, we will discuss the transport interaction rate defined
by $\Gamma _{trans}=n\; \sigma _{trans}$, where the transport cross
section $\sigma _{trans}=\int d\sigma \> (1-\cos \theta)$ enters the
collision term of the transport equation in the case of a plasma with
long range interactions, for which the interaction rate is dominated
by distant collisions \cite{r30}. Here $\theta $ denotes the
scattering angle in the center of mass system. Hence the mean free
path as well as the thermalization time may rather be given by the
inverse of the transport interaction rate \cite{r30}. The relevance
of this rate for the QGP has been pointed out in
Ref.\cite{r23,r31,r32}. The transport interaction rate is also
closely connected to the shear viscosity beyond the relaxation time
approximation \cite{r13,r33}.

The factor $1-\cos \theta $ cuts off small scattering angles and
therefore  small momentum transfers, too. Thus the transport
interaction rate is only logarithmically infrared divergent in naive
perturbation theory (Coulomb logarithm \cite{r30}), but finite
using the Braaten-Pisarski method since it is sensitive to the scale
$gT$ only. Hence dynamical screening of the magnetic interaction,
which is included in the effective gluon propagator, is sufficient
\cite{r34} leading to $\Gamma _{trans}\sim \alpha _s^2\, T\,
\ln(1/\alpha_s)$ \cite{r13,r29,r31}. Note the $\alpha _s^2$
dependence as expected from naive perturbation theory since there is
no enhancement by a factor $1/\alpha _s$ caused by a quadratic
infrared divergence. Also the logarthmic term stems from the
sensitivity to the scale $gT$ instead of $g^2T$ as it is the case for
$\Gamma $ \cite{r29}.

Up to now the transport interaction rates based on the
Braaten-Pisarski method have been calculated only within the leading
logarithm approximation \cite{r13} i.e., the coefficient of
$1/\alpha _s$ under the logarithm has not been determined. However,
if we want to extrapolate the result for the transport interaction
rate in the weak coupling limit to realistic values of $\alpha _s
=$ 0.2 -- 0.5, we should not neglect this coefficient.

One aim of the present paper is the calculation of the transport
interaction rate beyond the leading logarithm approximation using
the Braaten-Pisarski method. Furthermore we will discuss the
consequences of $\Gamma $ and $\Gamma _{trans}$ for realistic values
of $\alpha_s$. Finally, we will compare the results of the
Braaten-Pisarski method with the widely used approximation of the
naive perturbation theory regularizing the infrared singularities
simply by using the Debye screening mass as an infrared cut-off.

\section{Interaction Rates}

There are two equivalent ways of computing the interaction rates
either using matrix elements or self energies \cite{r35}. Considering
for example elastic quark-quark scattering we may find the
corresponding interaction rate $\Gamma_{qq}$ from the matrix element
via \cite{r8}
\begin{eqnarray}
\Gamma _{qq} & = & {1\over 2p}\> \int {d^3p' \over (2\pi )^3 2p'}\>
[1-n_F(p')]\> \int {d^3k\over (2\pi )^3 2k}\> n_F(k)\nonumber \\
&\times & \int {d^3k'\over (2\pi )^3 2k'}\> [1-n_F(k')]\> (2\pi )^4\>
\delta ^4(P+K-P'-K')\> 6N_f\> \langle |{\cal M}(qq\rightarrow qq)|^2
\rangle .\nonumber \\
\label{e1}
\end{eqnarray}
The four momentum of the incoming particle is denoted by $P=(p,{\bf
p})$, where quark masses are neglected and $p=|{\bf p}|$. (We
consider only the case of a QGP containing up and down quarks for
which the bare masses are negligible compared to the temperature of
the QGP.) The quark of the heat bath from which the quark under
consideration is scattered off has the momentum $K=(k,{\bf k})$.
Momenta with a prime belong to the outgoing particles. The
Fermi-Dirac distribution functions are given by $n_F(k)=1/[\exp
(k/T)+1]$, while $N_f$ denotes the number of thermalized flavors in
the QGP. The matrix element is averaged over the spin and color
degrees of freedom of the incoming particles and summed over the ones
of the final state. The factor $6N_f$ comes from summing over the
possible spin, color, and flavor states of the quark with momentum
$K$. The diagrams which enter the matrix element to lowest order are
shown in Fig.1a.

In the case of quark-antiquark scattering we have to replace the
diagrams of Fig.1a by the ones of Fig.1b. In the case of quark-gluon
scattering we have to substitute the Fermi distributions by the
Bose-Einstein distributions $n_B(k)=1/[\exp (k/T)-1]$, the Pauli
blocking factor $1-n_F(k')$ by the Bose enhancement factor
$1+n_B(k')$, and the factor $6N_f$ by 16 (number of gluonic degrees
of freedom). If we wish to consider the scattering of an incoming
gluon by the partons of the QGP we have to deal with the diagrams of
Fig.1c and 1d.

The second possibility to determine the quark interaction rate is
given by the imaginary part of the quark self energy on mass shell
\cite{r8}:
\begin{equation}
\Gamma _q(p)=-{1\over 2p}\> [1-n_F(p)]\> tr\left [\gamma ^\mu P_\mu
\, Im \Sigma (p,{\bf p})\right ].
\label{e2}
\end{equation}
The equivalence of the expressions (\ref{e1}) and (\ref{e2}) can be
seen from cutting the self energy of Fig.2a through the fermion
lines. Eq.(\ref{e2}) is the starting point for applying the
Braaten-Pisarski method by considering the quark self energy of
Fig.2b, where the effective gluon propagator contains the resummed
one-loop gluon self energy in the high temperature limit
\cite{r36,r37}. (Note that Fig.2b contains the diagram of Fig.2a if
the high temperature limit is used for the fermion loop, called hard
thermal loop approximation \cite{r18}, in the latter.) It is not
necessary to take an effective quark-gluon vertex into account
because the external quark with a momentum of the order of the
temperature is hard. Also because the interaction rate falls off
rapidly for large momentum transfers $q\equiv |{\bf p}-{\bf p}'|$,
$\Gamma \sim \int dq/q^3$, i.e., only small momentum transfers
contribute, a bare quark propagator is sufficient. An effective
quark propagator and quark-gluon vertex would contribute to higher
order in $\alpha _s$ only.

Since the final result for observables using the Braaten-Pisarski
method is gauge independent we are free of choosing any gauge. Using
Coulomb gauge and the approximation $p,k\simeq 3T\gg q\sim gT$ the
quark interaction can be written as \cite{r8}
\begin{equation}
\Gamma _q={C_Fg^2T\over 2\pi }\>\int _0^\infty dq\> q\> \int
_{-q}^{q}{d\omega \over \omega }\> \left [\rho _l(\omega ,q)
+\left (1-{\omega ^2\over q^2}\right )\> \rho _t(\omega ,q)
\right ],
\label{e3}
\end{equation}
where $C_F=4/3$ is the Casimir invariant of the fundamental
representation. The interaction rate of a hard gluon is simply
obtained by replacing $C_F$ by the Casimir invariant of the adjoint
representation $C_A=3$ \cite{r26}. The four momentum of the exchanged
gluon is denoted by $P-P'=K'-K\equiv Q=(\omega, {\bf q})$, and the
discontinuous parts of the longitudinal and transverse spectral
functions, related to the effective gluon propagator $\Delta _{l,t}$
through $\rho _{l,t}(\omega, q)=Im\Delta _{l,t}(\omega, q)/\pi$, are
given by ($-q\leq \omega \leq q$) \cite{r38}:
\begin{eqnarray}
\rho _l(\omega, q) & = & {3 m_g^2\omega \over 2q}\> \Biggl [ \left (
q^2+3m_g^2-{3m_g^2\omega \over 2q}\, \ln {q+\omega \over q-\omega}
\right )^2+\left ({3\pi m_g^2\omega \over 2q}\right )^2\Biggr ]^{-1},
\nonumber \\
\rho _t(\omega, q) & = & {3m_g^2\omega (q^2-\omega ^2)\over 4q^3}\>
\Biggl [\left (q^2-\omega ^2+{3m_g^2\omega^2\over 2q^2}\, \left
(1+{q^2-\omega^2\over 2\omega q}\, \ln {q+\omega \over q-\omega}
\right )\right )^2\nonumber \\ & + & \left ({3\pi m_g^2\omega
(q^2-\omega ^2)\over 4q^3}\right )^2\Biggr ]^{-1}, \nonumber \\
\label{e4}
\end{eqnarray}
where $m_g^2\equiv (1+N_f/6)\, g^2T^2/3$ may be interpreted as an
effective gluon mass generated by the interaction with the thermal
ensemble of the QGP. This expression was derived assuming the high
temperature approximation $\omega $, $q\ll T$.

The integration over $q$ can be performed analytically, whereas the
one over $\omega $ has to be done numerically. The result for the
longitudinal part of the interaction rate corresponding to the
exchange of a longitudinal gluon is given by \cite{r7}
\begin{equation}
\Gamma _q^l=1.098\> C_F\> \alpha _s\> T.
\label{e5}
\end{equation}
Note that the interaction rate is independent of the number of
flavors $N_f$ due to a cancellation of the factors $m_g^2$ in
(\ref{e4}) after integration. Although a larger number of thermal
flavors corresponds to an increase of the number of scattering
partners enlarging the interaction rate, the screening mass is also
increased cancelling this enlargement. Also the interaction rate does
not depend on the external momentum $p$ in the $p\gg \omega, q$
limit.

The transverse part of the interaction rate $\Gamma _q^t$, on the
other hand, is still infrared divergent, although the infrared
behavior has been improved from a quadratic singularity in the bare
two loop case (Fig.2a) to a logarithmic due to dynamical screening of
magnetic fields. On the other hand, there is no static magnetic
screening in the transverse part of the spectral function (or the
effective propagator) i.e., the denominator of $\rho _t$ vanishes in
the static limit $\omega =0$, $q\rightarrow 0$, while the denominator
of $\rho _l$ is given by $\mu _D^2 \equiv 3m_g^2$ in this limit. In
other words, the high temperature limit of the perturbatively
calculated gluon self energy contains static electric screening
(Debye mass $\mu _D$) but no static magnetic screening. In QCD,
however, static magnetic screening may arise non-perturbatively from
monopole configurations due to the gluon self interaction. Indeed,
lattice as well as semiclassical calculations show the excistence of
a magnetic screening mass, $m_{mag}^2\simeq 15\, \alpha _s^2\, T^2$
\cite{r39,r40}. Thus static magnetic screening is provided on the
momentum scale $g^2T$.

After all, in order to obtain an estimate of the transverse
interaction rate we consider the nearly static limit ($\omega \ll q$)
of (\ref{e4}) \cite{r25}:
\begin{eqnarray}
\rho _l(\omega \ll q) & = & {3m_g^2\omega \over 2q(q^2+3m_g^2)^2},
\nonumber \\ \rho _t(\omega \ll q) & = & {3m_g^2\omega q\over 4q^6
+(3\pi m_g^2 \omega /2)^2}.\nonumber \\
\label{e6}
\end{eqnarray}
Inserting (\ref{e6}) into (\ref{e3}) the integrations can be done
exactly, leading to
\begin{eqnarray}
\Gamma _q^l & = & C_F\> \alpha _s\> T,\nonumber \\
\Gamma _q^t & = & C_F\> \alpha _s\> T\> \ln {\kappa \over \alpha _s}.
\nonumber \\
\label{e7}
\end{eqnarray}
We observe that the nearly static limit is a good approximation
(within 10\%) for the longitudinal rate. The logarithmic term of the
transverse rate comes from assuming a hypothetical infrared cutoff of
the order $g^2T$. For example, the magnetic screening mass \cite{r25}
or the interaction rate itself i.e., the imaginary part of the quark
propagator \cite{r22,r23}, have been suggested as such an infrared
regulator. The coefficient $\kappa $ cannot be calculated using the
Braaten-Pisarski technique but must await the developement of
non-perturbative methods at finite temperature for dynamical
quantities \cite{r41}.

For a rough estimate we propose
\begin{equation}
\Gamma _q=\Gamma _q^l+\Gamma _q^t\simeq (2\pm 1)\> C_F \alpha _s\> T,
\label{e8}
\end{equation}
assuming $\alpha _s\simeq 0.3$ under the logarithm and $\kappa
{\buildrel <\over  \sim }2$. A justification for the latter
assumption may come from choosing $\mu _D$ as an upper limit and
$m_{mag}$ as a lower limit for the integration over $q$. The gluon
interaction rate, corresponding to the scattering processes shown
in Fig.1c and 1d, is given by $\Gamma _g= (C_A/C_F)\, \Gamma _q=(9/4)
\, \Gamma _q$. Because it is also independent of the number of
flavors, it holds for the QGP as well as for the pure gluon gas.

Compared to the real part of the dispersion relation for thermal
partons, $\omega \simeq k\simeq 3T$, the damping rate $\gamma
=\Gamma /2$ defined by the imaginary part of the self energy, is not
really small, $\gamma _g\simeq 3\alpha _sT$ for realistic values of
$\alpha _s$. This anomalously large damping \cite{r22,r23}
indicates that interactions in the QGP are important, at least for
temperatures not too far above the phase transition in accordance
with the recently emerging picture of the QGP \cite{r42}.

An alternative method to the Braaten-Pisarski method based on
(\ref{e2}) is given by inserting the $t$-channel diagrams of Fig.1
in the $-t\equiv -(P-P')^2\ll s\equiv (P+K)^2$ approximation into
(\ref{e1}), where the effective gluon propagator  $\Delta _{l,t}$ is
used instead of the bare one \cite{r43}. This method has been shown
to be equivalent to the self energy calculation (Eq.(\ref{e2}) to
(\ref{e4})) in the case of the energy loss of a massive fermion in a
hot plasma \cite{r8}.

Recently, Pisarski proposed an empirical way of including the
magnetic mass and an imaginary part of the fermion self energy in the
parton damping rates \cite{r44}. In the case of a hard, massless
quark, using the magnetic mass of \cite{r39,r40}, $\Gamma _q\simeq
0.8\, C_F\, \alpha _s\, T$ follows from his investigation assuming
$\alpha _s=0.3$ under the logarithm of the transverse part. This
result is smaller than the estimate (\ref{e8}), since the transverse
part of it turns out to be negative for realistic values of the
coupling constant.

Next we discuss a much simpler, widely used approximation (see, for
example, Ref.\cite{r4,r5,r6}) based on the naive perturbation theory
i.e., bare propagators and vertices, where the Debye mass $\mu _D^2=
3m_g^2$ is simply introduced by hand as an infrared regulator into
the gluon propagator. It should be noted that in this case the Debye
mass also cuts off the magnetic divergence without justification.
Then the interaction rate is easily calculated from
\begin{equation}
\Gamma =n\> \sigma =\int {d^3k\over (2\pi )^3}\> \rho (k)\> \int dt\>
{d\sigma \over dt},
\label{e9}
\end{equation}
where the parton momentum densities are given by $\rho _q(k)=12\,
N_f\, n_F(k)$ for quarks plus antiquarks and
$\rho _g(k)=16\, n_B(k)$ for gluons, respectively. The cross
sections in the small momentum transfer limit ($-t\ll s$) containing
the Debye mass read
\begin{equation}
{d\sigma \over dt}=\zeta \> {2\pi \alpha _s^2\over (t+\mu _D^2)^2},
\label{e10}
\end{equation}
where the color
factor $\zeta =4/9$ for quark-quark scattering, $\zeta =1$ for
quark-gluon scattering, and $\zeta =9/4$ for gluon-gluon scattering.
In contrast to the complete calculation (to leading order in the
coupling constant) using the Braaten-Pisarski method, the result
depends weakly on the number of flavors. In the case of two flavors
($N_f=2$) we find
\begin{eqnarray}
\Gamma _q(N_f=2) & \simeq & 1.1\> \alpha _s \> T,\nonumber \\
\Gamma _g(N_f=2) & \simeq & 2.5\> \alpha _s\> T \nonumber \\
\label{e11}
\end{eqnarray}
and in the pure gluonic case
\begin{equation}
\Gamma _g(N_f=0)\simeq 2.2\> \alpha _s\> T.
\label{e12}
\end{equation}
Comparing with (\ref{e8}) the simple approach leading to (\ref{e11})
or (\ref{e12}) appears to underestimate the interaction rates by
about a factor of two. However, it agrees with Pisarski's result
\cite{r44}.

Finally, we discuss the consequences of the present results for
typical values, $T=300$ MeV and $\alpha _s=0.3$, expected at RHIC
and LHC. Using (\ref{e8}) we arrive at relaxation times
\begin{eqnarray}
\tau _g & \simeq & (0.5\pm 0.3)\> fm/c,\nonumber \\
\tau _q & \simeq & (1.0\pm 0.5)\> fm/c \nonumber \\
\label{e13}
\end{eqnarray}
indicating a rapid thermalization of the QGP and the maintenance of
the local thermal equilibrium during the expansion phase by comparing
with a typical expansion time $\tau _{expan}>1$ fm/c \cite{r4}.
Similar results have been found from Monte Carlo simulations of
ultrarelativistic heavy ion collisions \cite{r2,r3,r4,r45,r46,r47}.
Furthermore the results (\ref{e13}) support the prediction of a
two-stage equilibration i.e., there is first a thermal equilibrium of
the gluonic component before a complete thermalization is achieved
later on because of the stronger interaction of the gluons compared
to the quarks \cite{r1,r4}.

\section{Transport Interaction Rates}

The relevant physical quantities, such as mean free path and
equilibration time, in a plasma with long range interactions are
described rather by the transport interaction rates than by the
ordinary ones \cite{r23,r30,r31,r32}. The transport rates are
obtained from the latter by introducing a weight $1-\cos \theta $
under the integrals of (\ref{e1}) or (\ref{e2}), defining $\Gamma
_{trans}\equiv \int d\Gamma\, (1-\cos \theta )$. Here $\theta $
denotes the scattering angle in the center of mass system:
$\cos \theta =({\bf p}\cdot {\bf p}')/(pp')=1+2t/s$. Thus the
transport factor $1-\cos \theta =-2t/s=2q^2\, (1-\omega ^2/q^2)/s$ is
proportional to the square of the momentum transfer $q^2$. This
additional factor $q^2$ changes the infrared and ultraviolet
behavior of the interaction rate completely. The transport
interaction rate behaves like $\Gamma _{trans}\sim \int dq/q$ in
naive perturbation theory. Therefore the soft as well as the hard
momentum transfer regimes contribute to $\Gamma _{trans}$. This is
very similar to the energy loss of a charged particle in a
relativistic plasma, where an additional factor $\omega ^2$ appears
compared to the interaction rate \cite{r7,r8}. Quantities which are
logarithmically infrared divergent in naive perturbation theory turn
out to be finite using the Braaten-Pisarski method for soft momentum
transfers (dynamical screening). Such quantities can be calculated
using the method proposed by Braaten and Yuan \cite{r48}. According
to this, introducing a separation scale $q^\star $, the soft and hard
contributions are calculated separately. For the soft contribution
($q<q^\star $) resummed propagators and vertices have to be used,
whereas bare Green's functions are sufficient for the hard one
($q>q^\star $). Assuming $gT\ll q^\star \ll T$ the otherwise
arbitrary scale $q^\star $ drops out at the end by adding the soft
and the hard contribution, reflecting the completeness of the
effective perturbation theory. In the following the Braaten-Yuan
method will be used for computing the transport interaction rate
following the example of the energy loss \cite{r8} as close as
possible.

We will present the calculation of the gluon transport rate in a pure
gluon plasma ($N_f=0$) in detail, quoting only the results for
the quark and gluon transport rates in a QGP of two active flavors
afterwards. The soft contribution follows from (\ref{e3}) introducing
the transport weight $1-\cos \theta $ under the integral and using
$q^\star $ as an upper limit for the $q$-integration:
\begin{equation}
\Gamma _{g,trans}^{soft}={C_Ag^2T\over \pi s}\>\int _0^{q^\star} dq\>
q^3\> \int _{-q}^{q} {d\omega \over \omega }\> \left (1-{\omega
^2\over q^2}\right )\> \left [\rho _l(\omega ,q)+\left (1-{\omega ^2
\over q^2}\right )\> \rho _t(\omega ,q)\right ].
\label{e14}
\end{equation}
The integral over $q$ can be done exactly, while the $\omega
$-integral has to be evaluated numerically. This has been done
already in Ref.\cite{r13} yielding ($m_g^2=4\pi \alpha _sT^2/3$)
\begin{eqnarray}
\Gamma _{g,trans}^{soft} & = & {3C_Ag^2T\over 2\pi s}\> m_g^2\>
\left [\ln \left ({{q^\star}^2 \over m_g^2}\right )-1.379\right ]
\nonumber \\ & = & {24\pi \alpha _s^2T^3\over s}\> \left [\ln \left
({{q^\star}^2\over \alpha _sT^2}\right )-2.811\right ].\nonumber \\
\label{e15}
\end{eqnarray}
In Ref.\cite{r13} the hard contribution has not been computed.
However, from general arguments ($q^\star $-cancellation) \cite{r48}
we know that the hard contribution has to be of the form $\Gamma
_{g,trans}^{hard}=B\, [\ln (T^2/{q^\star }^2)+A_{hard}]$, where
$B=24\pi \alpha _s^2T^3/s$ and the constant $A_{hard}$ has to be
determined from a detailed calculation of the hard contribution.
Thus, if we are only interested in a logarithmic accuracy (leading
logarithm approximation) valid in the weak coupling limit, we end up
with \cite{r13}
\begin{equation}
\Gamma _{g,trans}={24\pi \alpha _s^2T^3\over s}\> \ln {1\over \alpha
_s}.
\label{e16}
\end{equation}
However, since the strong coupling constant is not small for
realistic values expected in ultrarelativistic heavy ion collisions,
the knowledge of the coefficient under the logarithm is essential.
Therefore the constant $A_{hard}$ of the hard contribution has to be
determined. Before we turn to this, let us note that the $\ln
(1/\alpha _s)$-term in (\ref{e16}) arises from the sensitivity of
$\Gamma _{trans}$ to the momentum scale $gT$, while the logarithmic
term in the interaction rate (\ref{e7}) comes from a sensitivity to
the scale $g^2T$ \cite{r29}. Thus the transport interaction rate can
be calculated completely to leading order in the coupling constant
using the Braaten-Pisarski method in contrast to $\Gamma $. These
entirely different properties of the two different kinds of rates
originate, of course, from the additional factor $q^2$ in
$\Gamma _{trans}$.

For the hard contribution ($q>q^\star $) it is sufficient to use the
bare gluon propagator. However, in contrast to the soft contribution
the $-t\ll s$ approximation does not hold any more and the $u$- and
$s$-channel diagrams of Fig.1 cannot be neglected. The corresponding
amplitudes have been calculated a long time ago in Feynman gauge
\cite{r49,r50}. Since the on-shell matrix elements are gauge
invariant, regardless of the momenta integrated over, there is no
problem in adopting those results although the soft contribution has
been evaluated in Coulomb gauge \cite{r8,r51}.

However, a new difficulty arises now from the divergence of the
$u$-channel contribution for $u=(P-K')^2\rightarrow 0$ i.e.,
$-t\rightarrow s$. (For massless particles $s+t+u=0$ holds.) The
$u$-channel divergence can be regulated in the same way as the
$t$-channel singularity, if we choose the transport factor $(\sin
\theta )^2/2=2tu/s^2$ instead of $1-\cos \theta $. This choice is
justified because the transport weight $1-\cos \theta $ has been
introduced only for small scattering angles \cite{r30} for which it
is identical to $(\sin \theta )^2/2$. The latter also restores the
$t$-$u$-channel symmetry in the transport cross sections of
quark-quark and gluon-gluon scattering. Furthermore, the shear
viscosity coefficient beyond the relaxation time approximation also
contains a factor $\sin ^2 \theta $ \cite{r13,r16,r32,r33}. Finally,
parton collisions with scattering angles near $0^\circ $ as well as
$180^\circ$ are less important for achieving an isotropic momentum
distribution (thermal equilibrium) indicating the physical
significance of a transport rate defined by a weight proportional to
$\sin ^2 \theta $ instead of $1-\cos \theta $ for the equilibration
process.

For calculating the hard contribution of the transport interaction
rate in a gluon gas we start from (\ref{e1}) modified to gluon-gluon
scattering and including the factor $(\sin \theta )^2/2$:
\begin{eqnarray}
& \Gamma & _{g,trans}(N_f=0)={1\over 2p}\> \int {d^3p' \over (2\pi
)^3 2p'}\>[1+n_B(p')]\> \int {d^3k\over (2\pi )^3 2k}\> n_B(k)
\nonumber \\
& \times & \int {d^3k'\over (2\pi )^3 2k'}\> [1+n_B(k')]\> (2\pi )^4
\> \delta ^4(P+K-P'-K')\> 16\> \langle |{\cal M}(gg\rightarrow gg)|
^2\rangle \>{\sin ^2\theta \over 2},\nonumber \\
\label{e17}
\end{eqnarray}
where the matrix element contains the scattering diagrams of Fig.1d.
While the hard contribution of the energy loss of a heavy fermion
with mass $M\gg T$ could be calculated exactly \cite{r8,r9}, this is
not possible for (\ref{e17}). Therefore we propose the following
approximations: First we assume $n_B(p')\simeq n_B(p)$ and $n_B(k')
\simeq n_B(k)$. These simplifications hold as long as $q=|{\bf p}-
{\bf p}'|=|{\bf k}'-{\bf k}|$ is not too large or $-t$ is not of the
order of $s$. This assumption may be justified because the transport
factor $(\sin \theta )^2/2$ cuts off those momenta effectively. While
we will neglect $n_B(p)$ because of $\langle p\rangle \simeq 3T$, we
do not set $n_B(k)=0$. As a matter of fact, the Bose enhancement
factor $1+n_B(k)$ is important for the exact matching of the soft and
hard parts i.e., for the cancellation of $q^\star $.

Using the definition of the differential cross section \cite{r52}
(\ref{e17}) may now be written as
\begin{equation}
\Gamma _{g,trans}(N_f=0)=\int {d^3k\over (2\pi )^3}\> 16\, n_B(k)\,
[1+n_B(k)]\> \int dt\> \left ({d\sigma \over dt}\right )_{gg}\> {2tu
\over s^2}.
\label{e18}
\end{equation}
The $k$-integration over the Bose distribution functions gives a
factor $8T^3/3$, compared to $16 \zeta (3)T^3/\pi ^2=1.95\, T^3$
neglecting the Bose enhancement factor. The differential cross
section for $gg\rightarrow gg$ scattering according to the diagrams
of Fig.1d are taken from Ref.\cite{r49}
\begin{equation}
\left ({d\sigma \over dt}\right )_{gg}={9g^4\over 64\pi s^2}\> \left
(-{us\over t^2}-{st\over u^2}-{tu\over s^2}+3\right ),
\label{e19}
\end{equation}
where a factor $1/2$ has been included to account for the identical
particles in the final state \cite{r29}. The hard contribution
follows from (\ref{e18}) by restricting the $t$-integration from $-s$
to ${-q^\star }^2$. Assuming $s\gg {q^\star }^2$, (\ref{e18})
together with (\ref{e19}) results in
\begin{equation}
\Gamma _{g,trans}^{hard}(N_f=0)= {24\pi \alpha _s^2T^3\over s}\>
\left [\ln \left ({T^2\over {q^\star }^2}\right )+\ln \left ({s\over
T^2} \right )-{19\over 15}\right ].
\label{e20}
\end{equation}
In order to proceed with the calculation we replace the Mandelstam
variable $s$ under the logarithm by its average thermal value
$\langle s\rangle= 2\langle p\rangle\, \langle k\rangle =14.59\, T^2$
for gluon momenta $\langle p\rangle =\langle k\rangle=2.701\, T$.
Then we arrive at
\begin{equation}
\Gamma _{g,trans}^{hard}(N_f=0)={24\pi \alpha _s^2T^3\over s}\>
\left [\ln \left ({T^2\over {q^\star }^2}\right )+1.414\right ].
\label{e21}
\end{equation}
Adding up the soft and hard contributions, (\ref{e15}) and
(\ref{e21}), the separation scale $q^\star $ drops out as required:
\begin{eqnarray}
\Gamma _{g,trans}(N_f=0) & = & {24\pi \alpha _s^2T^3\over s}\>
\left [\ln \left ({1\over \alpha _s}\right )-1.397\right ]\nonumber
\\ & = & {24\pi \alpha _s^2T^3\over s}\> \ln {0.25\over \alpha _s}
\nonumber \\ & \simeq & 5.2\, \alpha _s^2\, T\> \ln {0.25\over \alpha
_s},\nonumber \\
\label{e22}
\end{eqnarray}
where we have used $s\simeq \langle s\rangle $ in the last equation.

The result (\ref{e22}) may be compared to the simple approximation of
using bare propagators with the Debye mass as infrared regulator.
Since the degree of the infrared singularity is only logarithmic this
amounts approximately to using (\ref{e18}), where the separation
scale ${q^\star }^2$ is replaced by $\mu _D^2=4\pi \alpha _s T^2$ in
the upper limit of the $t$-integration. In this way we find
\begin{equation}
\Gamma _{g,trans}\simeq {24\pi \alpha _s^2T^3\over s}\> \ln {0.33
\over \alpha _s}.
\label{e23}
\end{equation}
Comparing to (\ref{e22}), one realizes that the -- to leading order
in $\alpha _s$ exact -- result (\ref{e22}) may be obtained by using
an effective infrared cut-off of $1.32\, \mu _D^2$ instead of $\mu
_D^2$.

In the case of a QGP with two flavors we have to consider the
diagrams of Fig.1a-c in addition. Modifying (\ref{e18}) to these
processes, the corresponding calculations are a little bit more
involved than in the purely gluonic case. For example, we have to be
careful about the flavors of the final state e.g., the flavors of the
final state quarks of the $u$- and $s$-channel diagrams of Fig.1a and
1b have to be identical. Furthermore there is no $t$-$u$-channel
symmetry in the quark-gluon scattering process. After all the
$u$-channel singularity is cancelled by the transport factor $(\sin
\theta )^2/2$ because it is only logarithmic in this case, rendering
the use of an effective quark propagator unnecessary. Also the gluon
"mass" $m_g$ depends now on $N_f$ and, finally, in the case of quarks
(or antiquarks) with momenta $K$ and $K'$ we have to replace in
(\ref{e18}) the factor 16 by $6N_f$, $n_B(k)$ by $n_F(k)$, and
$1+n_B(k)$ by $1-n_F(k)$. The soft contribution for the quark rate
follows from (\ref{e14}) by replacing $C_A$ by $C_F$. Putting
everything together we end up with
\begin{eqnarray}
\Gamma _{g,trans}(N_f=2) & = & \left (1+{N_f\over 6}\right ) \>
{24\pi \alpha _sT^3\over s'}\> \ln {0.17\over \alpha _s}\nonumber \\
& \simeq & 6.6\, \alpha _s^2\, T\> \ln {0.17\over \alpha _s},
\nonumber \\ \Gamma _{q,trans}(N_f=2) & = & \left (1+{N_f\over 6}
\right )\> {32\pi \alpha _sT^3\over 3s''}\> \ln {0.14\over \alpha _s}
\nonumber \\ & \simeq & 2.5\, \alpha _s^2\, T\> \ln {0.14\over \alpha
_s}, \nonumber \\
\label{e24}
\end{eqnarray}
where $s'=(1+N_f/6)/(1/s_{gg}+N_f/6s_{gq})$ and $s''=(1+N_f/6)
/(1/s_{gq}+N_f/6s_{qq})$. Here $s_{gg}$ is the Mandelstam variable
for two gluons, $s_{gq}$ for one gluon and one quark, and $s_{qq}$
for two quarks in the initial state, leading to $\langle s'\rangle =
15.13\, T^2$ and $\langle s''\rangle = 17.65\, T^2$.

Applying the transport interaction rates (\ref{e22}) and (\ref{e24})
to ultrarelativistic heavy ion collisions we encounter a serious
problem: The results (\ref{e22}) and (\ref{e24}), obtained to lowest
order perturbation theory, become negative for realistic values of
$\alpha _s>0.2$. Hence we cannot draw any conclusions regarding
thermalization times at RHIC and LHC from (\ref{e24}). However, the
validity of the Landau collision integral containing the transport
cross section beyond the logarithmic approximation depends on the
condition that the characteristic length $l$ over which the
distribution function varies significantly must be large compared
with the screening length $1/\mu _D$ i.e., $l\, \mu _D\gg 1$
\cite{r30}, which is not fulfilled for $1/l\simeq T$ and $\mu _D
\simeq 4 {\sqrt \alpha _s} T$. Thus the physical significance of the
transport interaction rate is somewhat obscure in the QGP.

The unphysical, negative results for the transport rates for
realistic values of $\alpha _s$ arise from the separate calculation
of the soft and hard contributions, which works only in the weak
coupling limit $g\ll 1$. In the soft and hard contributions the
assumption $T\gg q^\star \gg gT$ is essential for achieving the
cancellation of $q^\star $. Of course, this assumption cannot be
fulfilled for $g{\buildrel >\over \sim }1$ rendering (\ref{e15})
and (\ref{e21}) negative, although each contribution is positive by
itself without any restriction on $q^\star $, since they are
equivalent to integrals over squares of magnitudes of matrix
elements. A similar problem occured in (\ref{e23}) by using $\mu _D$
as an upper limit for the $t$-integration instead of a regulator in
the gluon propagator and assuming $s\gg \mu _D$, which does not hold
for $g{\buildrel >\over \sim }1$ any more.

The problem of an unphysical, negative result already appeared in
the computation of the energy loss of a heavy quark in the QGP
\cite{r9}. There it was argued that by using the effective
perturbation theory for the entire momentum range this problem may be
circumvented, increasing, however, the complexity of the calculation
drastically. It is not possible simply to use the effective gluon
propagator for the whole integration range, since it is well defined
only for $\omega $, $q\ll T$. Gauge invariance and completeness in
the order of the coupling constant then demand the use of effective
vertices and quark propagators in addition, corresponding to
including higher orders in $g$ \cite{r9,r48}. Thus, in order to
guarantee positive results and gauge invariance at the same time for
values of the coupling constant $g{\buildrel >\over \sim }1$, one has
to go beyond the lowest order in $g$. In the weak coupling limit, on
the other hand, the lowest order contribution, here $\alpha _s^2\,
\ln (1/\alpha _s)$, is sufficient. Furthermore such a unphysical
behavior for realistic values of the coupling constant has also been
observed in the calculation of the gluon plasma frequency beyond
leading order \cite{r53}. In all of these cases, a negative result
shows up, if $g$ exceeds a critical value of about 1.

Finally, we will discuss the implications of the transport
interaction rates for the viscosity of the QGP
\cite{r13,r16,r32,r43,r54,r55,r56}. In Ref.\cite{r13,r16} the shear
viscosity coefficient $\eta $ has been calculated from
\begin{equation}
\eta _i\simeq {4\over 15}\> {\epsilon _i\over \Gamma _i^{\eta}},
\label{e25}
\end{equation}
where $i=q,g$, $\epsilon _i$ is the energy density of the quarks and
gluons in the QGP, and $\Gamma _i^\eta =2\, \Gamma _{i,trans}$
\cite{r13}. (The factor of two comes from using the weight $\sin
^2\theta $ instead of $(\sin \theta )^2/2$ in the definition of the
shear viscosity coefficient \cite{r33}.) Inserting (\ref{e24}) into
(\ref{e25}) we find
\begin{equation}
\eta =\eta _g+\eta _q={T^3\over \alpha _s^2}\> \left [{0.11\over \ln
(0.17/\alpha _s )}+{0.37\over \ln (0.14/\alpha _s)}\right ].
\label{e26}
\end{equation}
This should be compared to the most elaborate calculation of $\eta $
(albeit in the leading logarithm approximation) using a variational
method for the Boltzmann equation resulting in $\eta =1.16\,
T^3/[\alpha _s^2\ln (1/\alpha _s)]$ \cite{r43}. For $\alpha _s<0.1$
both results are comparable.

\section{Conclusions}

Before presenting our conclusions, we discuss the validity of the
approximations used here. The distinction between soft and hard
momenta and the cancellation of the separation scale $q^\star $,
which are the foundations of the Braaten-Pisarski and the Braaten-Yuan
method \cite{r18,r48}, rely on the weak coupling limit assumption
$g\ll 1$. In contrast, realistic values of $\alpha _s>0.2$ imply
$g>1.5$. Since $\alpha _s$ is expected to decrease only
logarithmically with increasing temperature, $g\ll 1$ is not even
fulfilled at the extreme temperature of the Planck scale. On the
other hand, the Braaten-Pisarski method is nothing but an improvement
of the usual perturbation theory at finite temperature which should
work at temperatures above twice the critical according to
comparisions with lattice QCD \cite{r57,r58}. Furthermore, comparing
the effective gluon "mass" $m_g$, calculated using the
non-perturbative Hartree approximation, with perturbative results
shows a difference between the both results by less than 30\% at
$g=1.5$ \cite{r41}. Also in the case of the energy loss of an
energetic quark in the QGP it has been shown that the result depends
only weakly on the assumption $gT\ll q^\star \ll T$ \cite{r11}. Those
observations indicate that the assumption $g\ll 1$ should be merely
regarded as a mathematical trick and not as a physical restriction.
Therefore we believe that the Braaten-Pisarski method not only
provides a consistent treatment of QCD at high temperatures taking
into account at the same time important physics as collective effects
of the non-ideal relativistic plasma e.g., screening, but also gives
results for realistic situations which are correct within about a
factor of 2 \cite{r59}, as long as logarithmic factors $\ln (const/
\alpha _s)$ are not too close to zero or negative as it is the case
for the transport interaction rate.

We have estimated the ordinary interaction rate of thermal quarks and
gluons by using the effective perturbation theory of Braaten and
Pisarski, for which the use of an effective gluon propagator is
sufficient in the case of thermal partons. Due to the missing static
magnetic screening in the transverse part of the effective gluon
propagator and the absence of an imaginary part of the quark
propagator we still encounter a logarithmic infrared singularity.
Assuming a reasonable cut-off, a rough estimate has been obtained,
$\Gamma _g=(6.0\pm 3.0)\, \alpha _s\, T$ for gluons and $\Gamma _q
=(2.7\pm 1.3)\, \alpha _s\, T$ for quarks, which corresponds to
relaxation times of the order $\tau = (0.5\pm 0.3)$ fm/c for gluons
and $\tau = (1.0\pm 0.5)$ fm/c for quarks. This indicates a rapid
thermalization of the gluon component (two-stage equilibration
\cite{r1}) and a maintenance of the local thermal equilibrium during
the expansion phase of the possibly formed QGP at RHIC and LHC in
accordance with computer simulations of ultrarelativistic heavy ion
collisions \cite{r2,r3,r45}.

On the other hand, in a plasma with long range interactions as in QCD
the physically relevant quantity should be the transport rather than
the ordinary interaction rate. The transport rate follows from the
ordinary one by introducing a transport weight containing the
scattering angle in the center of mass system. Due to this factor the
infrared behavior of the interaction rate is completely changed. The
transport rate turns out to be finite using the Braaten-Pisarski
method because dynamical screening suffices now. We have calculated
the transport interaction rate for thermal quarks and gluons
beyond the leading logarithm approximation by
decomposing it into soft and hard parts according to the prescription
of Braaten and Yuan \cite{r48}. The soft contribution has been
computed by using the effective gluon propagator of the
Braaten-Pisarski method, while the hard contribution has been treated
using bare propagators and vertices.

Compared to the ordinary interaction rate the transport rate is
reduced by a factor of $\alpha _s$ caused by the improved infrared
behavior due to the transport weight. For a QGP of two active flavors
$\Gamma _{g,trans} \simeq 6.6\, \alpha _s^2\, T\, \ln
(0.17/\alpha _s)$ for gluons and $\Gamma _{q,trans}\simeq 2.5\,
\alpha _s^2\, T\, \ln (0.14/\alpha _s)$ for quarks have been found.
The surprisingly small values of the coefficients under the logarithm
show that $\Gamma _{trans}$ is only meaningful for $\alpha _s
{\buildrel <\over \sim }0.1$. Improving the calculation by using the
Braaten-Pisarski method over the entire momentum range increases the
complexity of the calculation enormously, including higher orders of
$g$. Hence no statement about the consequences (e.g., thermalization
times, mean free paths) for realistic values of the coupling constant
can be given here. For this purpose, at least a calculation beyond
the lowest order perturbation theory is required. However, the
transport interaction rate obtained here suggests that it may be much
smaller than the ordinary rate. Thus the realization of a local
thermal equilibrium in relativistic heavy ion collisions seems to be
questionable assuming the transport rates to be responsible for
thermalization, in contrast to computer simulations. However, neither
in HIJING \cite{r2} nor in the parton cascade \cite{r3} transport
cross sections for the fundamental parton interactions are used.

Furthermore the shear viscosity, which is proportional to the inverse
of the transport interaction rate \cite{r13}, has been obtained for
the first time beyond the leading logarithm approximation. For values
of $\alpha _s<0.1$, for which the result is well defined, it is large
and comparable to the ones obtained by using the leading logarithm
approximation \cite{r13,r16,r43}. This supports the idea that
dissipation cannot be neglected in hydrodynamic descriptions of the
expansion phase of the QGP in ultrarelativistic heavy ion collisions
\cite{r16}.

Finally, we have compared the rates obtained from the
Braaten-Pisarski method, which are complete to leading order in the
coupling constant, with the widely used approach of using bare
propagators including the Debye mass as an infrared regulator. While
the latter approximation works well for transport rates and energy
losses \cite{r8}, it seems to underestimate the ordinary rates. This
observation suggests that the use of the Debye regulator is justified
for quantities which are logarithmically infrared divergent in naive
perturbation theory as the transport rates or the energy loss, but
might be questionable for quadratically infrared divergent quantities
as the ordinary interaction rate.

\acknowledgments
The author would like to thank T.S. Bir\'o for valuable discussions.
The work was supported by the BMFT and GSI Darmstadt.

\begin{figure}
\caption{Lowest order Feynman diagrams for $qq\rightarrow qq$ (a),
$q\bar q\rightarrow q\bar q$ (b), $qg\rightarrow qg$ (c), and
$gg\rightarrow gg$ (d) scattering.}
\end{figure}

\begin{figure}
\caption{Lowest order quark self energy contributions to the quark
interaction rate using naive perturbation theory (a) and using the
Braaten-Pisarski method (b).}
\end{figure}


\begin{references}
\bibitem{r1}
E. Shuryak, Phys. Rev. Lett. {\bf 68}, 3270 (1992).

\bibitem{r2}
X. N. Wang and M. Gyulassy, Phys. Rev. D {\bf 44}, 3501 (1991).

\bibitem{r3}
K. Geiger and B. M\"uller, Nucl. Phys. {\bf B369}, 600 (1992).

\bibitem{r4}
T. S. Bir\'o, E. van Doorn, B. M\"uller, M. H. Thoma, and X. N.
Wang, Duke Univ. Report No. DUKE-TH-93-46, 1993 (to be published in
Phys. Rev. C).

\bibitem{r5}
J. D. Bjorken, Fermilab Report No. PUB-82/59-THY, 1982 (unpublished).

\bibitem{r6}
B. Svetitsky, Phys. Rev. D {\bf 37}, 2484 (1988).


\bibitem{r7}
M. H. Thoma and M. Gyulassy, Nucl. Phys. {\bf B351}, 491 (1991).

\bibitem{r8}
E. Braaten and M. H. Thoma, Phys. Rev. D {\bf 44}, 1298 (1991).

\bibitem{r9}
E. Braaten and M. H. Thoma, Phys. Rev. D {\bf 44}, R2625 (1991).

\bibitem{r10}
S. Mr\'owczy\'nski, Phys. Lett. B {\bf 269}, 383 (1991).

\bibitem{r11}
M. H. Thoma, Phys. Lett. B {\bf 273}, 128 (1991).

\bibitem{r12}
Y. Koike and T. Matsui, Phys. Rev. D {\bf 45}, 3237 (1992).

\bibitem{r13}
M. H. Thoma, Phys. Lett. B {\bf 269}, 144 (1991).

\bibitem{r14}
M. Gyulassy and M. Pl\"umer, Phys. Lett. B {\bf 243}, 432 (1990).

\bibitem{r15} M. Gyulassy, M. Pl\"umer, M. H. Thoma, and X. N. Wang,
Nucl. Phys. {\bf A538}, 37c (1992).

\bibitem{r16}
P. Danielewicz and M. Gyulassy, Phys. Rev. D {\bf 31}, 53
(1985).

\bibitem{r17}
See, for example, U. Heinz, K. Kajantie, and T. Toimela,
Ann. Phys. (N.Y.) {\bf 176}, 218 (1987).

\bibitem{r18}
E. Braaten and R. D. Pisarski, Nucl. Phys. {\bf B337}, 569 (1990).

\bibitem{r19}
R. Baier, G. Kunstatter, and D. Schiff, Phys. Rev. D {\bf 45}, 4381
(1992).

\bibitem{r20}
H. Nakkagawa, A. Ni\'egawa, and B. Pire, Phys. Lett. B {\bf 294}, 396
(1992).

\bibitem{r21}
E. Braaten and R. D. Pisarski, Phys. Rev. D {\bf 46}, 1829 (1992).

\bibitem{r22}
T. Altherr, E. Petitgirad, and T. del Ri\'o Gaztelurrutia,
Phys. Rev. D {\bf 47}, 703 (1993).

\bibitem{r23}
V. V. Lebedev and A. V. Smilga, Ann. Phys. (N.Y.) {\bf 202}, 229
(1990).

\bibitem{r24}
V. V. Lebedev and A. V. Smilga, Phys. Lett. B {\bf 253}, 231 (1991).

\bibitem{r25}
R. D. Pisarski, Phys. Rev. Lett. {\bf 63}, 1129 (1989).

\bibitem{r26}
E. Braaten, Nucl. Phys. B (Proc. Suppl.) {\bf 23B}, 351 (1991).

\bibitem{r27}
C. P. Burgess and A. L. Marini, Phys. Rev. D {\bf 45}, 17 (1992).

\bibitem{r28}
A. Rebhan, Phys. Rev. D {\bf 46}, 482 (1992).

\bibitem{r29}
H. Heiselberg and C. J. Pethick, Phys. Rev. D {\bf 47}, 769 (1993).

\bibitem{r30}
E. M. Lifshitz and Pitaevskii, {\it Physical Kinetics} (Pergamon,
New York, 1981).

\bibitem{r31}
H. Heiselberg, G. Baym, C. J. Pethick, and J. Popp, Nucl. Phys.
{\bf A544}, 569c (1992).

\bibitem{r32}
S. Mr\'owczy\'nski, in {\it Quark-Gluon Plasma}, edited by R. Hwa
(World Scientific, Singapore, 1990).

\bibitem{r33}
F. Reif, {\it Fundamentals of Statistical and Thermal Physics}
(McGraw-Hill, New York, 1965).

\bibitem{r34}
C. J. Pethick, G. Baym, and H. Monien, Nucl. Phys. {\bf A498}, 313c
(1989).

\bibitem{r35}
H. A. Weldon, Phys. Rev. D {\bf 28}, 2007 (1983).

\bibitem{r36}
V. V. Klimov, Zh. Eksp. Teor. Fiz. {\bf 82}, 336 (1982) [Sov. Phys.
JETP {\bf 55}, 199 (1982)].

\bibitem{r37}
H. A. Weldon, Phys. Rev. D {\bf 26}, 1394 (1982).

\bibitem{r38}
R. D. Pisarski, Physica {\bf A158}, 146 (1989).

\bibitem{r39}
T. A. de Grand and D. Toussaint, Phys. Rev. D {\bf 25}, 526 (1989).

\bibitem{r40}
T. S. Bir\'o and B. M\"uller, Duke Univ. Report No. DUKE-TH-92-42,
1992 (to be published in Nucl. Phys. A).

\bibitem{r41}
M. H. Thoma, Mod. Phys. Lett. {\bf A7}, 153 (1992).

\bibitem{r42}
See, for example, T. S. Bir\'o, Int. Jour. Mod. Phys. {\bf 1}, 39
(1992).

\bibitem{r43}
G. Baym, H. Monien, C. J. Pethick, and D. G. Ravenhall, Phys. Rev.
Lett. {\bf 64}, 1867 (1990).

\bibitem{r44}
R. D. Pisarski, Brookhaven Report No. BNL-P-1/92, 1992 (unpublished).

\bibitem{r45}
K. Geiger, Phys. Rev. D {\bf 46}, 4965, 4986 (1992).

\bibitem{r46}
B. M\"uller and X. N. Wang, Phys. Rev. Lett. {\bf 68}, 39 (1992).

\bibitem{r47}
T. S. Bir\'o, B. M\"uller, and X. N. Wang, Phys. Lett. B {\bf 283},
171 (1992).

\bibitem{r48}
E. Braaten and T. C. Yuan, Phys. Rev. Lett. {\bf 66}, 2183 (1991).

\bibitem{r49}
B. L. Combridge, J. Kripfganz, and J. Ranft, Phys. Lett. {\bf 70B},
234 (1977).

\bibitem{r50}
R. Cutler and D. Sivers, Phys. Rev. D {\bf 17}, 196 (1978).

\bibitem{r51}
H. A. Weldon (private communication).

\bibitem{r52}
J. D. Bjorken and S. D. Drell, {\it Relativistic Quantum Mechanics}
(McGraw-Hill, New York, 1964).

\bibitem{r53}
H. Schulz, Hannover Univ. Report No. ITP-UH 8/93, 1993 (unpublished).

\bibitem{r54}
A. Hosoya and K. Kajantie, Nucl. Phys. {\bf B250}, 666 (1985).

\bibitem{r55}
R. Horseley and W. Schoenmaker, Nucl. Phys. {\bf B280}, 768 (1987).

\bibitem{r56}
S. V. Ilyin, O. A. Mogilevski, S. S. Smolyansky, and G. M. Zinovjev,
Phys. Lett. B {\bf 296}, 385 (1992).

\bibitem{r57}
See, for example, B. Petersen, Nucl. Phys. {\bf A525}, 237c (1991).

\bibitem{r58}
M. Gao, Phys. Rev. D {\bf 41}, 626 (1990).

\bibitem{r59}
J. Kapusta, P. Lichard, and D. Seibert, Phys. Rev. D {\bf 44}, 2774
(1991).

\end{references}
\end{document}